\documentclass[10pt]{iopart}
\newcommand{\BEQ}{\begin{equation}}
\newcommand{\EEQ}{\end{equation}}
\newcommand{\BEA}{\begin{eqnarray}}
\newcommand{\EEA}{\end{eqnarray}}
\renewcommand{\H}{{\cal {H}}}

\newcommand{\p}{\partial}

\newcommand{\I}{{\cal {I}}}
\newcommand{\nn}{\nonumber }
\newcommand{\Kt}{{\tilde K}}
\newcommand{\Ht}{{\tilde H}}

\newcommand{\bam}{{\overline m}}

\newcommand{\bamu}{{\overline {\mu}}_2}
\newcommand{\tinf}{{\mbox{\hspace*{-0.2 mm}\tiny $\infty$}}}

\begin{document}
\title{Exactly Solvable Model Glass with a Facilitated Dynamics}
\author{Luca Leuzzi and Theo M Nieuwenhuizen}
\address{Universiteit van Amsterdam,
 Valckenierstraat 65, 1018 XE Amsterdam, The Netherlands}
\ead{leuzzi@science.uva.nl}

\begin{abstract}
A model glass with fast and slow processes is studied.
The statics is simple and the facilitated slow dynamics is
exactly solvable. 
The main features of a fragile glass take place: Kauzmann transition,
Vogel-Fulcher law,  Adam-Gibbs relation and aging.
The time evolution
can be so slow that a quasi-equilibrium occur at a time dependent
effective temperature. The same effective temperature is derived from
the Fluctuation-Dissipation ratio, which supports the applicability
of out of equilibrium thermodynamics.
\end{abstract}
\submitto{\JPCM}
\pacs{}
\maketitle
\section{Introduction}
\label{intro}
We   present  the outcome of our
investigation of  an exactly
 solvable  model glass that shows all of the basic features 
of much more complicated real glasses, such as 
aging \cite{BBM,BCKM},
 diverging relaxation time to equilibrium (Vogel-Fulcher-Tammann-Hesse
(VFTH) like \cite{VFTH}), configurational entropy satisfying the Adam-Gibbs
 relation \cite{AG}, 
Kauzmann transition \cite{KAUZ} and 
violation of Fluctuation-Dissipation Theorem \cite{CK93,BCKM}.

In studying the model we  make use of a particular parallel Monte Carlo 
   dynamics
that retains the fundamental  collective nature  of the glassy 
dynamics
 and that can be carried out analytically on our model. 
We will see how this dynamics undergoes a huge slowing down as the 
system is cooled down and which kind of 
 aging dynamics the system sets out. 
We can implement such dynamics even below the
Kauzmann temperature, thus getting information in a regime where few
 analytic results are known. 

The main motivations for this work are twofold.
Firstly we want to get more insight in the glassy dynamics, 
in its various aspects,
exploiting the  analytical solvability of our model. Indeed,  
every detected feature of the glassy behaviour
can be connected in a direct correspondence
with given elements of the model, thanks  to its simplicity.
We can even switch on and off certain properties or certain 
dynamic behaviours, tuning the model parameters or implementing
the facilitated dynamics in alternative ways.
Furthermore, the configurational entropy
is exactly computable
(see section \ref{sec2}) as a function of the 
dynamic variables of the model.

The second goal is to check the generality of the concept of
{\em effective temperature}, very often discussed in literature in
many different approaches (see for instance in
\cite{BCKM,CK93,TOOL,SKT,CR,N00,N2,FV,fictive}) 
and to see whether the possibility exists of inserting 
such a parameter in the construction of a consistent out of 
equilibrium thermodynamic theory.
Even though the physics of our model is
 simple, we shall find general aspects of the results
by formulating them in the thermodynamic language (see section \ref{sec5}). 

In glasses, 
the exponential divergence of  time-scales
(as opposed to  the algebraic divergence in  standard
continuous phase transitions), might 
 induce an asymptotic decoupling of the time-decades \cite{ANGELL}.
Exploiting this, we build the dynamics of our model on
the reasonable assumption that, in a glassy
system that has aged a long time $t$,  all processes
with equilibration time much less than $t$ (the
$\beta$ processes) are in equilibrium  while those evolving on 
 time scales much larger than $t$ are still
quenched, leaving the processes with time scale of order $t$ 
(i.e., the $\alpha$ processes), as the only interesting ones. 
To model this, 
we will introduce by hand two different kind of variables evolving
on two very different (decoupled) time-scales.

Such an asymptotic decoupling of time scales, that is the input for the
present  model and for a set of other models of the same class 
\cite{N00,BPR,NPRL98,LN2,GR},
 could be the basis for 
a generalization of equilibrium thermodynamics to systems 
out of equilibrium \cite{N00}.

In section \ref{sec2} we introduce the model and in section \ref{sec3}
the facilitated dynamics that we apply to it.
The dynamical behaviour of the one-time observables will be shown
in section \ref{sec4}. In section \ref{sec5} an out of equilibrium 
formulation of thermodynamics is proposed.
From the study of the two-time variables dynamics (in section \ref{sec6})
we show that, quite reasonably since we are not at equilibrium,
 the Fluctuation-Dissipation Theorem is not  valid anymore.
We will see in section \ref{sec6} that from the Fluctuation-Dissipation Ratio
we can derive an independent definition of effective temperature.




 \section{ The model and its static properties }\label{sec2}

\subsection{Hamiltonian and constraint on the configuration space}
The model displaying a fragile glass behaviour was
introduced in \cite{NCM} and widely studied in
\cite{LN1}. It is
 described by the following local Hamiltonian:
\BEQ
{\cal{H}}[\{x_i\},\{S_i\}]=\frac{1}{2} K
\sum_{i=1}^{N}x_i^2
-H\sum_{i=1}^{N}x_i
-J\sum_{i=1}^{N}x_i S_i
-L\sum_{i=1}^{N}S_i
\label{Hmodel} \ ,
\EEQ

\noindent where $N$ is the size of the system and 
$\{x_i\}$ and $\{S_i\}$ are continuous variables, 
the last satisfying a spherical constraint: $\sum_i S_i^2=N$.
We will  call them respectively harmonic oscillators
and spherical spins.
$K$ is the Hooke elastic constant, $H$ is an external field acting on
the harmonic oscillators, $J$ is the coupling constant between $\{x_i\}$ 
and $\{S_i\}$ on the same site $i$,
 and $L$ is the external field acting on the spherical spins.
 A separation
of time scales
 is introduced by hand: the spins represent the fast modes and  the
harmonic oscillators the slow ones.
We  assume that the $\{S_i\}$ relax to equilibrium
 on a 
time scale  much shorter
than the one of the harmonic oscillators.
From the point of view of the motion of the $\{x_i\}$, the spins are just
a noise. 
To describe the long time regime of the $\{x_i\}$,
 we  can average over this noise by performing the computation 
of the $\{S_i\}$ partition function,
obtaining an effective Hamiltonian depending only on the $\{x_i\}$,
that  determines the dynamics of these variables .
 Using the 
saddle point approximation for large $N$  we find:
\BEQ
\fl Z_S(\{x_i\})=
\int\left(\prod_{i=1}^{N} d S_i\right)
 \exp\left\{-\beta{\cal H}\left[\{x_i\},\{S_i\}\right]\right\}\,\,
\delta\left(\sum_{i=1}^{N}S_i^2-N\right) 
\simeq
e^{-\beta{\cal{H}}_{\rm eff}(\{x_i\})} 
\label{Zs}
\EEQ
where we introduce  the effective Hamiltonian
${\cal{H}}_{\rm eff}(\{x_i\})\equiv -T\log Z_S(\{x_i\})$, that is the free
energy for a given configuration of $\{x_i\}$, as
\BEQ               
\fl {\cal{H}}_{\rm eff}(\{x_i\})=
\frac{K}{2}  m_2 N - H  m_1 N - w N+
 \frac{TN}{2}\log\frac{w+\frac{T}{2}}{T} \ .
\label{Heff}
\EEQ                            
We defined the short-hands
\BEQ
\fl m_1\equiv\frac{1}{N}\sum_{i=1}^N x_i \hspace*{0.4cm}\ , \hspace*{0.4 cm}
m_2\equiv\frac{1}{N}\sum_{i=1}^N x_i^2 \hspace*{0.4 cm}\ , \hspace*{0.4 cm} 
w\equiv\sqrt{J^2 m_2+2J L m_1+ L^2+ \frac{T^2}{4}} \ .
\label{abbrev_m}
\EEQ        
                                 
The effective Hamiltonian, equation (\ref{Heff}),
 can also be written  in terms of the internal energy $U(\{x_i\})$
and of the entropy $S_{\rm ep}(\{x_i\})$
of the equilibrium processes (i.e. the spins):
\BEA
{\cal{H}}_{\rm eff}(\{x_i\})=U(\{x_i\})-T S_{\rm e p}(\{x_i\})  \ ;
\label{def:Heff}\\
\fl U(\{x_i\})= \frac{K}{2}  m_2 N - H m_1 N - w N+ \frac{TN}{2}
\hspace*{0.2 cm},\hspace*{0.2cm}
S_{\rm ep}(\{x_i\})=\frac{N}{2}-\frac{N}{2}\log \frac{w+T/2}{T} \ .
\label{USep}
\EEA

The  model  is also characterized  by 
a constraint on the phase space, introduced 
 to avoid the existence of the single 
global minimum, and 
implementing a large degeneracy of the allowable lowest states. 
The constraint is taken on the $\{x_i\}$,
 thus concerning the long time regime. It reads:
\BEQ
m_2-m_1^2\geq m_0 \ ,
\label{CONSTRAINT}
\EEQ
where $m_0$ is a model parameter. 
It is a  fixed, but arbitrary, strictly positive constant.
This constraint, applied to the harmonic oscillator dynamics, is
a way to reproduce the behavior of good glass formers, 
i.e., substances for which nucleation of the crystal phase
is especially unlikely even at very slow cooling rates
(e.g., network formers $B_2O_3$ and  $Si O_2$, molecular organics
such as glycerol and  atactic polystyrene, and  different multicomponent
 liquid mixtures). 
These are  substances for which there are 
 non-crystalline packing modes for the particles composing them, 
that have intrinsically low energy. The amorphous 
configurations are thus favored.
 In general the crystal state still exists, at
lower energy, but the probability of nucleating a crystal 
instead of a glass is  negligible.
 In specific cases 
(binary solutions) the glassy state can even be lower in energy 
than the crystalline one and is thermodynamically stable with 
respect to any crystal configuration \cite{KA}.

In the next section
we will impose 
a Monte Carlo dynamics \cite{BPR,BPPR}
satisfying this constraint and coupling 
the otherwise non-interacting $\{x_i\}$ in a dynamic way.
We will study such dynamics analytically.

To shorten the notation 
 we define  the modified ``spring constant'' 
${\tilde{K}}$ and ``external field'' ${\tilde{H}}$:
\BEQ
{\tilde{K}}=K-\frac{J^2}{w+T/2},\hspace*{2 cm}
{\tilde{H}}=H+\frac{J L}{w+T/2}
\label{def:tilde}
\EEQ
We stress  that ${\tilde{K}}$ and ${\tilde{H}}$ are actually functions of
the $\{x_i\}$ themselves (through $m_1$ and $m_2$, occurring  in $w$).


\subsection{Statics at heat-bath temperature $T$}
Before describing the facilitated dynamics employed we sketch
very briefly the static of the model.
The partition function of the whole system at equilibrium is:
\BEA
\fl Z(T)&=&\int {\cal{D}}x {\cal{D}} S \exp\left[-\beta{\cal{H}}(\{x_i\},\{S_i\})\right] \delta\left(\sum_i x_i^2-N\right)=
\label{Z_static}
\\
\nn
\fl &=&
\int d m_1 d m_2 \exp\left\{-\beta N\left[\frac{K}{2} m_2 -H m_1-w+\frac{T}{2}\log\left(\frac{w+T/2}{T}\right)-\frac{T}{2}\left(1+\log(m_2-m_1^2)\right)\right]\right\}
\EEA
The new object (with respect to equation (\ref{Zs}))
 appearing  in the exponent of the above expression 
is the  configurational entropy
\BEQ
{\cal{I}}\equiv\frac{N}{2}\left[1+\log(m_2-m_1^2)\right] \ .
\label{I}
\EEQ

The saddle point equations 
are found minimizing 
the expression between square brackets in (\ref{Z_static})
with respect to $m_1$ and $m_2$.
This yields
\BEQ
\frac{\Ht(\bam_1,\bam_2)}{\Kt(\bam_1,\bam_2)}=\bam_1 \hspace*{1cm} ,
\hspace*{1 cm}
\bam_2-\bam_2^2=\frac{T}{\Kt(\bam_1,\bam_2)} \ .
\label{bam}
\EEQ
The form of the solutions $\bam_1(T)$, $\bam_2(T)$
 is quite complicated because each of these
equations is actually a fourth order equation, but they can 
be explicitly computed.


\section{Facilitated Monte Carlo dynamics}
\label{sec3}
We assume as  dynamics a 
parallel Monte Carlo (MC) dynamics for the harmonic oscillators. 
This kind of analytic Monte Carlo approach was first introduced in
\cite{BPPR}, and later applied to the exactly solvable
harmonic oscillator model~\cite{BPR}
(which is just our model after setting $J=L=0$)
and to  the  spherical spin model~\cite{NPRL98,N00}
(which is the present model after setting $H=K=0$ and considering the 
$\{x_i\}$ as quenched random variables).
 The  dynamical
model thus obtained
with a very simple Hamiltonian and a contrived dynamics has the
benefit of being even
solvable analytically, which yields a much deeper insight into its
properties than numerical simulations. 
Moreover, in the long-time domain the dynamics looks
quite reasonable in regard to what one might expect of any system with a
VFTH-law in its approach to equilibrium.

In a Monte Carlo step a random updating of the variables is performed
($x_i\to x'_i=x_i+r_i/\sqrt{N}$) where the $\{r_i\}$ have a Gaussian 
distribution with zero mean and variance $\Delta^2$.
We define $x\equiv{\cal{H}}(\{x'_i\})-{\cal{H}}(\{x_i\})$ as
the energy difference 
between the new and the old state.
If $x>0$ the move is 
 accepted with a probability $W(\beta x)\equiv
\exp(-\beta x)$; else 
 it is  always accepted ($W(\beta x)=1$).
The updating is made in parallel.
It is the parallel nature of the updating that allows
 the collective behavior leading to exponentially divergent time scales
in  models with no  interactions between particles such us ours.
A sequential updating would not produce any glassy effect.
 This dynamics may induce glassy behavior in situations 
where ordinary Glauber dynamics \cite{GLAUBER} would not. 
In our model the parallel  dynamics mimics the presence of interactions
between atoms in realistic glasses, where a large internal cooperativeness 
occur.
In this respect the constraint is not essential.
A parallel MC dynamics that does not take into account the constraint on the 
$\{x_i\}$ still maintains glassy behaviour (see \cite{LN2} for the study
of such dynamical version of our model).
Here, however, we will only look at the dynamics with a built-in constraint.
For different examples of dynamics implying non trivial collective behavior
the reader can look, for instance, at the  $n$-spin facilitated kinetic 
Ising model \cite{FA84,FR} or at the kinetic lattice-gas model \cite{KA,KPS}.

In this section we will  show the basic steps leading to dynamical equations.
 For further details refer to
 \cite{N00,BPR,LN2,LN1}.
In a Monte Carlo step the quantities $\sum_ix_i=Nm_1$ and 
$\sum_ix_i^2=Nm_2$ are updated. We denote their change by 
$y_1$ and $y_2$, respectively. Their  distribution function is, 
for given values   of $m_1$ and $m_2$,
\BEA
\fl p(y_1,y_2|m_1,m_2)&\equiv &\int\prod_i \frac{d r_i}{\sqrt{2\pi\Delta^2}}
e^{-{r_i^2}/({2\Delta^2})}\,\,
\delta\left(\sum_i {x'}_i-\sum_ix_i-y_1\right)\,
\delta\left(\sum_i {x'}_i^{2}-\sum_ix_i^2-y_2\right)
\nn \\ &=&
\frac{1}{4\pi \Delta^2\sqrt{m_2-m_1^2}}
\exp\left(-\frac{y_1^2}{2\Delta^2}-\frac{(y_2-\Delta^2-2y_1m_1)^2}
{8\Delta^2(m_2-m_1^2)}\right) \ .
\label{Ptrans}
\EEA
Neglecting the variations of $m_1$ and $m_2$ 
 of order $\Delta^2/N$ we can express the energy difference as
\BEQ
x=\frac{\Kt}{2} \ y_2 -{\tilde{H}} \ y_1,
\label{x}
\EEQ

In terms of $x$
and $y=y_1$ the distribution function can be formally written
 as the product of two  Gaussian  distributions:
\BEA
\fl p(y_1,y_2|m_1,m_2)d y_1 d y_2
	&=&\hspace*{ 6 mm}d x\hspace*{7 mm}p(x|m_1,m_2)
\hspace*{ 1.3 cm} d y\hspace*{5 mm}p(y|x,m_1,m_2)
\label{PROBDIST}\\
&=&\frac{dx}{\sqrt{2\pi \Delta_x}}
	\exp\left(-\frac{(x-{\overline{x}})^2}{2 \Delta_x}\right)
	\frac{ dy}{\sqrt{2\pi \Delta_y}}
	\exp\left(-\frac{(y-{\overline{y}}(x))^2}{2 \Delta_y}\right)
	\nn
\EEA

\noindent where 
\BEA
\fl {\overline{x}}=\Delta^2{\tilde{K}}/2, \hspace*{ 2.5cm}
\Delta_x=\Delta^2{\tilde{K}}^2
(m_2-m_1^2)+\Delta^2{\tilde{K}}^2\left(m_1-{\tilde{H}}/{\tilde{K}}\right)^2,\\
\fl  {\overline{y}}(x)=\frac{m_1-{\tilde{H}}/{\tilde{K}}}{
m_2-m_1^2+\left(m_1-{\tilde{H}}/{\tilde{K}}\right)^2}
\frac{x-{\overline{x}}}{\tilde{K}},
  \hspace*{0.2 cm}
\Delta_y=\frac{\Delta^2(m_2-m_1^2)}
{m_2-m_1^2+\left(m_1-{\tilde{H}}/{\tilde{K}}\right)^2}.
\label{def:MCave}
\EEA

To represent a fragile glass the dynamics that 
we apply to the system is a generalization of the
analytic treatment of
 Monte Carlo dynamics introduced
 in \cite{BPPR}.
Also in this generalized case  the  dynamical
model with  a contrived dynamics can be
 analytically solved. 

We let $\Delta^2$, the variance 
 of the random updating $\{r_i\}$,
  depend on the distance from the 
constraint, i.e. on  the whole $\{x_i\}$ configuration before the 
Monte Carlo update:

\BEQ
\Delta^2(t)\equiv 8[m_2(t)-m_1^2(t)]
\left[\frac{B}{m_2(t)-m_1^2(t)-m_0}\right]^\gamma
\label{Delta}
\EEQ
where $B$, $m_0$ and $\gamma$ are constants.
The exponent $\gamma$ also enter the VFTH relaxation law, as we will see later 
on. In literature it is usually set equal to $1$, 
and an argument for this choice 
was given by  Adam and Gibbs\cite{AG}. 
An exact explanation for it  was provided by Kirkpatrick, Thirumalai and 
Wolynes in \cite{KTW89} and a further quantitative analysis is also 
reported in \cite{XW00}. 
However, their studies do not exclude 
 exponents $\gamma > 1$, always compatible with data,
 merely affecting the width of the fitting interval.  Analytic  
approaches \cite{KW87,ParisiVF} give  $\gamma=2$ in three dimensions.
Here we consider $\gamma$ as a model parameter, which can be
chosen below, equal to,  or above unity, and 
investigate  aspects of this standard picture.

The VFTH law that we obtain is a direct consequence of the special choice
(\ref{Delta}) for the MC update.
In the harmonic oscillator model and in the spherical spin model 
studied in 
\cite{N00,BPR,NPRL98}, the dynamics was performed
within this approach, but at fixed $\Delta$.
Both cases showed a relaxation time diverging 
at low temperature with an Arrhenius law, typical of {\em strong} glasses.
The same is found  by setting $m_0=0$ and $\gamma=1$ in the present
 model \cite{LN2}
but here we want, instead, to develop a model representing a 
{\em{fragile}} glass with a Kauzmann transition at a finite temperature.

The question whether detailed balance  is satisfied 
is also non-trivial  in our model. Indeed, 
it happens to be satisfied for this kind of dynamics
 only for large $N$. For exact detailed balance we should have 
\BEQ
p(x|m_1,m_2)\exp(-\beta x)=p(-x|m_1,m_2)
\EEQ
\noindent
but now, when we perform the inverse move $\{x'_i\}\to\{x_i\}$,
the probability distribution  also depends on the $\{r_i\}$
through $\Delta^2$ as defined in equation (\ref{Delta}). 
It can be verified that the violation is of order $1/N$.

The Monte Carlo  equations for the dynamics of $m_1$ and $m_2$  derived
from this construction  read 
\BEA
\fl \dot{m}_1\hspace*{-1 mm}=\hspace*{-1 mm}\int dy_1dy_2 W(\beta x) y_1  p(y_1,y_2|m_1,m_2) \hspace*{-1 mm}=\hspace*{-1 mm}
\int dx W(\beta x)  {\overline{y}}(x)  p(x|m_1,m_2)\ ,
\label{eq:m1}\\
\fl \dot{m}_2\hspace*{-1 mm}=\hspace*{-1 mm}\int dy_1dy_2  W(\beta x) y_2  p(y_1,y_2|m_1,m_2) \hspace*{-1 mm}=\hspace*{-1 mm}
\frac{2}{\tilde{K}}\hspace*{-1 mm}\int dx W(\beta x) 
 (x+{\tilde{H}}  {\overline{y}}(x)) p(x|m_1,m_2) 
\label{eq:m2}
\EEA

\section{Single-time dynamical observables}
\label{sec4}

The  dynamics of the system
can be  expressed in terms of   two combinations of
 $m_1$ and $m_2$. 
The first one, defined  as 
\BEQ
\mu_1\equiv\frac{\tilde{H}}{\tilde{K}}-m_1.
\label{defMU1}
\EEQ
\noindent  represents the distance from
the instantaneous equilibrium 
state.
 By instantaneous equilibrium state we mean that $\Ht$ and $\Kt$
 depend on the  values of $m_1$ and $m_2$ at a given time $t$.
For $t\to \infty$, at the true equilibrium, one has $\mu_1=0$.

The second dynamical variable  is defined as the distance from 
the constraint (\ref{CONSTRAINT}):
\BEQ
\mu_2\equiv m_2-m_1^2-m_0.
\EEQ
When $\mu_2=0$ the constraint is reached. This will happen if the temperature
is low enough ($T\leq T_0$) and the time large enough. 
$T_0$ is the highest temperature at which
the constraint is asymptotically ($t\to\infty$) reached by the system,
it is identified with the Kauzmann temperature \cite{LN1}.

The nearer the system goes to the constraint (i.e. the smaller the 
value of $m_2-m_1^2-m_0$), the larger the variance $\Delta^2$, 
implying almost always a refusal of the proposed updating.
In this way, in the neighborhood of the constraint,  the dynamics 
is very slow  and goes on through  very seldom but very
large moves, which  can be interpreted as activated processes.
When the constraint is reached the variance 
$\Delta^2$ becomes infinite and the system
 dynamics gets stuck.
The system does not evolve anymore towards equilibrium
 but it is blocked in one  single ergodic component
of the configuration space. 

In terms of $\mu_1$ and $\mu_2$, 
the equations of motion 
(\ref{eq:m1}),(\ref{eq:m2}) become
\BEA
\fl\dot{\mu_1}=-J Q \hspace*{-1mm}\int dx  \ W(\beta x)\  x \  p(x|m_1,m_2) 
\hspace*{-1mm}-\hspace*{-1mm}(1+ Q D) \hspace*{-1mm}\int dx \ W(\beta x)\  {\overline{y}}(x)\   p(x|m_1,m_2)
\label{mu1dot},
\\
\fl\dot{\mu_2}=\frac{2}{\tilde{K}}\int dx\  W(\beta x)\  x   p(x|m_1,m_2) 
+2 \mu_1  \int dx \ W(\beta x)\  {\overline{y}}(x) \  p(x|m_1,m_2)
\label{mu2dot},
\EEA
\noindent where $D$ and $Q$ are given by
\BEQ
\fl D\equiv H J+K L={\tilde{H}}J+{\tilde{K}}L \hspace*{1cm} , 
\hspace*{1 cm} Q\equiv \frac{J^2 D}{\Kt^3 w\left(w+T/2\right)^2}\ .
\label{DQ}
\EEQ

Above $T_0$ ordinary equilibrium will be achieved without reaching 
the constraint. 
The temperature is, then,  too high for the system to notice that there is a 
constraint at all on the  configurations (we are speaking about the 
asymptotic time regime), and this implies 
\BEQ 
\lim_{t\to \infty}\mu_2(t)= \bamu(T)=\frac{T}{\Kt_{\tinf}(T)}-m_0>0 \ ,
\label{bamu2}
\EEQ
\noindent where 
\BEQ
\Kt_{\tinf}(T)\equiv \lim_{t \to\infty}\Kt\left(m_1(t),m_2(t);T\right)=
\Kt\left(\bam_1(T),\bam_2(T)\right)  .
\EEQ
$\bam_1$ and $\bam_2$ are the solutions of the static self-consistent 
equations (\ref{bam}), if $T\geq T_0$.

For $T<T_0$ the second equation in (\ref{bam}) should be replaced by
\BEQ
\bam_2-\bam_1^2=m_0\hspace*{2 cm} \forall T<T_0\ .
\EEQ
Below $T_0$ the system goes to configurations arbitrarily
close to the constraint, and then stay there arbitrarily long.
By definition of $T_0$, we can write
\BEQ
m_0=\frac{T_0}{\Kt_{\tinf}(T_0)}\ .
\label{m0T0}
\EEQ

The equations of motion \ref{mu1dot} and \ref{mu2dot} can be solved in the 
long time regime,
 for fixed parameters (aging setup).
We notice that the value of  the VFTH exponent $\gamma$ discriminates
between different dynamic regimes if $\gamma>1$, 
$\gamma=1$ or $0<\gamma<1$ ~\cite{LN1} (the situation $\gamma=1$ remains
 model dependent even in the long time limit).
We find, to the leading orders of approximation for large times,
the following behavior for $\mu_2$ \cite{LN1}:
\BEQ
\mu_2(t)\simeq \frac{B}{\left[\log (t/t_0)+
c\log\left(\log(t/t_0)\right)\right]^{1/\gamma}} \ .
\label{MU2}
\EEQ

For $T \geq T_0$ the parameters $c$ and $t_0$ are
\BEQ
c=\frac{1}{2} \ ;
\hspace*{2 cm}
t_0\equiv \frac{\sqrt{\pi}(1+Q_{\tinf}D)}{8\gamma(1+P_{\tinf}+Q_{\tinf}D)} \ .
\EEQ
\noindent $t_0$ is of $O(1)$ for a large range of parameters values and 
$Q_{\tinf}$ is given by equation (\ref{DQ}) computed at $m_1=\bam_1(T)$ and 
$m_2=\bam_2(T)$.
The other function of temperature appearing in the above expression,
$P_{\tinf}$,  is defined as the infinite time limit of
\BEQ
P\equiv \frac{J^4 (m_2-m_1^2)}{2 \Kt w \left( w+T/2\right)^2} \ .
\label{P}
\EEQ
That means $P$ computed at $m_1=\bam_1(T)$ and 
$m_2=\bam_2(T)$.
The solution (\ref{MU2}) is valid in the aging regime, when
$t_0\ll t\ll\tau_{eq}(T)$.

Below $T_0$ the qualitative behavior of $\mu_2(t)$ 
 is the same, but $T$ is never reached. 
Concerning  the solution (\ref{MU2}) the only difference is in the values
\BEQ
\fl c=\frac{2+\gamma}{2\gamma} \ \ ;\hspace*{2 cm}
t_0\equiv \frac{B \sqrt{\pi}}{8\gamma}
\frac{\left(2 \Kt_{\tinf}(T) m_0-T\right)}{m_0^2\Kt_{\tinf}(T)
\left(\Kt_{\tinf}(T) m_0- T\right)} \ .
\EEQ

The dynamical behavior of $\mu_1$ depends not only on the temperature 
(above or below $T_0$) but also  on $\gamma$ being greater, equal to or less
 than one.
With respect to the relative weight of $\mu_1$ and 
$\mu_2$ we can identify different regimes\cite{LN1}. 
In this presentation we just state that  for  the regime
with  $T \geq T_0$ and  for the one with  $T < T_0$  and $\gamma>1$,
  $\mu_1(t)\ll\mu_2(t)$  and
a unique effective thermodynamic parameter
 can be properly defined in various
independent ways (see next section).

What we said up to now concerns the aging regime, but, when time 
grows on even larger time scales, finally approaching equilibrium
in the temperature regime $T\geq T_0$,
the equations of motion for any one-time observable $o(t)$
(magnetization,  energy, distance from the constraint, etc.) take the form
\BEQ
\dot o(t)\simeq -\frac{o(t)}{\tau_{eq}}
\EEQ
 From the study of the dynamics for very large times \cite{LN1,NCM}
we get a characteristic relaxation time to equilibrium
that depends on temperature following a generalized VFTH law:
\BEQ
\tau_{eq}\sim \exp\left(\frac{A}{T-T_0}\right)^{\gamma} \ . 
\EEQ
\noindent 
When $t\sim\tau_{eq}(T)\sim \exp\left[A/(T-T_0)\right]^{\gamma}$
 the ``distance'' $\mu_2$ becomes,
\BEQ
\bamu(T)\simeq\frac{B}
{\left[\left(\frac{A_{\rm f}}{T-T_0}\right)^{1/\gamma}\right]^{\gamma}}
\propto T-T_0 \ .
\EEQ

The parameter $T_0$ in the VFTH law is identified with the Kauzmann 
temperature, i.e. the temperature such that $\I(T_0)\equiv\I_0$ 
is the minimum
of the configurational entropy [and for any $T<T_0$ remains 
$\I(T)=\I_0$].
Moreover,  the specific heat displays a discontinuity at $T_0$:
at that temperature 
the model undergoes  a real thermodynamic phase transition.
The Adam-Gibbs relation
between relaxation time and configurational entropy density 
(equation (\ref{I})) is also achieved,
in the form\cite{LN1}
\BEQ
\tau_{\rm eq}\sim\exp\left(\frac{N}{\I-\I_0}\right)^{\gamma}\ .
\EEQ

\section{Effective temperature and out of equilibrium thermodynamics}
\label{sec5}

In this section we introduce
  effective parameters
in order to rephrase the dynamics of the system out of equilibrium
into a thermodynamic description (for a complete derivation see \cite{N00}). 

In \cite{LN1} we got, through different methods,  various  expressions
for the effective temperature $T_e$  as  function of the 
interaction parameters of the model and of the time evolution
of its observables. All of them were coinciding in the regime for $T>T_0$ and
$\gamma>1$.

 We want to shortly  recall one  particular  derivation of $T_e$.
 Knowing the solution of the dynamics at a given time $t$
a quasi-static approach 
can be followed by  computing the 
 partition function $Z_e$ of all the macroscopically 
equivalent states 
at the time  $t$.
In order to generalize the equilibrium thermodynamics we 
assume an effective temperature $T_e$ and an effective field $H_e$,
and substitute the Boltzmann-Gibbs equilibrium measure by 
$\exp (-\H_{\rm eff}(\{x_i\},T,H_e)/T_e)$, 
where $\H_{\rm eff}$ is given in (\ref{def:Heff})
 and the true external field $H$ has been substituted by
the effective field $H_e$.
As  we get the expression of the ``thermodynamic''
potential $F_e\equiv-T_e \log Z_e$  as a function
of macroscopic variables $m_{1,2}$ and effective parameters,
we can determine $T_e$ and $H_e$ 
minimizing $F_e$ with
respect to $m_1$ and $m_2$ 
and evaluating
the resulting analytic expressions at $m_{1,2}=m_{1,2}(t)$.

The partition function of the macroscopically equivalent states is:
\BEQ
\fl Z_e\equiv \int {\cal{D}}x 
\ \exp\left[-\frac{1}{T_e}\H_{\rm eff}(\{x_i\},T,H_e)
\right] \ 
\delta( N m_1-\sum_ix_i)\ \delta( N m_2-\sum_ix_i^2) \ .
\label{Ze}
\EEQ

From this  we  build the effective thermodynamic potential
as a function of $T_e$ and $H_e$, besides of $T$ and $H$, where
the effective parameters depend on time through the
time dependent values of $m_1$ and $m_2$, solutions of the dynamics:
\BEQ
 F_e(t)=U
-T{S_{\rm ep}}-T_e(t)\I
+[H-H_e(t)]N m_1(t), \label{Fe}
\EEQ
\noindent with 
\BEA
T_e(t)&=&\Kt\left(m_1(t),m_2(t)\right) \left[m_0+\mu_2(t)\right],
\label{Te}
\\
 H_e(t)& =&H-\Kt\left(m_1(t),m_2(t)\right) \mu_1(t) \ .
\label{He}
\EEA
 $T_e$ and $H_e$ are actually
a way of describing the evolution in time of the system out of equilibrium.
$U$ is the internal energy of the whole system,
${S_{\rm ep}}$ is the entropy  of the fast or equilibrium 
processes (the spherical spins) (see equation (\ref{USep}))
while ${\cal{I}}$ is the 
entropy of the slow, ''configurational'', processes
(the harmonic oscillators, see equation (\ref{I})). 
 The last term of $F_e$ replaces the $-HNm_1$ occurring
in $U$  by $-H_eNm_1$.
$U$, ${S_{\rm ep}}$ and $\I$ 
are 'state' functions, in the sense that  they depend on the state
described by $T$, $T_e$, $H$ and, if needed,  $H_e$.
 In the case where  only one  relevant  effective parameter $T_e$ remains,
 these functions do not depend on the path along which its
value has been reached.

As we already mentioned, for our VFTH relaxing model at $T\geq T_0$, and at
$T <T_0$ with $\gamma>1$, 
the effective temperature alone is
enough for a complete thermodynamic description of the dominant
physical phenomena ($H_e=H$).

\section{Two-time  variables: breaking of time-translation invariance 
and the fluctuation-dissipation relation}
\label{sec6}

In this section we compute the  correlation
and response functions which, unlike the energy and the quantities
 $m_1(t)$ and $m_2(t)$,
depend in a 
non-trivial way on two times, when the system is out of equilibrium,
 thus showing directly the loss of time translation invariance 
with respect 
to the case at equilibrium.
The aim of computing such quantities 
is also to build a Fluctuation Dissipation relation and look
at the meaning of the Fluctuation-Dissipation Ratio (FDR), 
$\p_{t'}C(t,t')/G(t,t')$,  far from equilibrium, and to compare
it with the effective temperature derived in other ways, e.g., 
as in (\ref{Te}).

The correlation functions between the thermodynamic
fluctuation of a quantity $m_a(t)$ at time $t$
and that of a quantity $m_b(t')$ at a different time $t'$
are defined as
\BEQ
C_{a b}(t,t')\equiv N \left<\delta m_a(t)\delta m_b(t')\right>, 
\ \ \ \ a, b=1, 2
\EEQ
\noindent where $\left< .... \right>$ is the average over 
the dynamic processes, i.e., the harmonic oscillators.

The response of an observable $m_a$ at time $t$ to a perturbation in 
a conjugate field $H_b$ at some previous time $t'$ takes the form
\BEQ
G_{a b}(t,t')\equiv\frac{\delta\left<m_a(t)\right>}{\delta H_b(t')},
\ \ \ \ a, b=1, 2
\EEQ
\noindent In our model $H_1=H$ and $H_2$=$-K/2$.

In order to be concise, in the following  we will  only give results without 
derivation; moreover, we will concentrate on the FDR for 
fluctuations of $\sum_i x_i=N m_1$, coupled to the external field
$H$, i.e., we only consider $a=b=1$.
For the complete derivation refer to \cite{LN1}.

Knowing
 the evolution of the two-time observables we can 
generalize the Fluctuation Dissipation Theorem
 defining another effective temperature, $T_e^{FD}$,
by means of the ratio between the derivative with respect to the 
initial time (also called  ``waiting'' time) $t'$ of the
correlation function $C_{11}$ and the response function $G_{11}$:
\BEQ
T_e^{F D}(t,t')\equiv\frac{\partial_{t'}C_{11}(t,t')}{G_{11}(t,t')}.
\EEQ

Dynamics varies strongly if $T$ is above or below the Kauzmann temperature and 
this difference produces different equations for the leading terms in
the correlation and response functions. Therefore, we present results for the 
two cases separately.

\subsection{High temperature case: $T>T_0$, $\forall \gamma$}

First of all we define
the time evolution function for the considered  time-scale sector as
\BEQ
{\tilde{h}}(\tau)\equiv\exp\left(-\int_0^{\tau}{\tilde{f}}(t)d t\right)
\label{htilde}
\EEQ
and the acceptance rate of the MC dynamics
\BEQ
\Upsilon=\frac{\exp\left(-\Gamma\right)}{\sqrt{\pi \Gamma}}
\frac{T_e}{2T_e-T}\ .
\EEQ
$T_e$ is, in  the above expression, just an abbreviation for 
$\Kt \left(m_0+\mu_2\right)$, as given in equation (\ref{Te}).
We will look at its relation with $T_e^{FD}(t,t')$.
The function ${\tilde{f}}$ in equation (\ref{htilde})
is, in this regime, 
\BEQ
\fl {\tilde{f}} = -4\Upsilon\left\{(1+Q D)\Gamma-\left[1+Q D-\frac{2 D Q P}{1+Q D}-\frac{D P(1+Q D)}{\gamma(1+P+Q D)}\right]
+O\left(\frac{1}{\Gamma}\right)\right\}
\label{ft}
\EEQ
\noindent with $D$, $Q$ and $P$  given respectively by 
(\ref{DQ}) and (\ref{P}).

The correlation function comes out to be
\BEQ
C_{11}(t,t')=C_{11}(t',t')\frac{\tilde{h}(t')}{\tilde{h}(t)}+
O(\mu_2^{1+\gamma}\Upsilon)\ ,
\EEQ
\noindent with 
\BEQ
C_{11}(t',t')\simeq 
\frac{m_0+\mu_2(t')}{1+Q(t') D} \ .
\EEQ

Following the approach of 
\cite{N00} we also  derive the response function:

\BEQ
 G_{11}(t,t')=G_{11}(t',t')\frac{\tilde{h}(t')}{\tilde{h}(t)}+
O(\mu_2^{1+\gamma}\Upsilon)\ ,
\EEQ
\noindent with 
\BEA
 G_{11}(t',t')&=&-\beta\int dy_1dy_2 W'(\beta x) y_1^2 p(y_1,y_2|m_1,m_2)
\\ \nonumber
 &=&
-\beta\int dx W'(\beta x)\Delta_y p(x|m_1,m_2)+O(\mu_2^2 \Upsilon)
\\
 &=&
\frac{4\Upsilon(t')\Gamma(t')}{\Kt(t')}-\frac{2\Upsilon(t')}{\Kt(t')}
+O(\mu_2 \Upsilon)\ .
\EEA

Eventually we get
\BEQ
T_e^{F D}(t,t')\simeq T_e(t')
\left[1+O\left(\mu_2(t')^\gamma\right)
+O\left(\mu_2(t')^2\right)
\right]\ .
\label{TeFD1}
\EEQ
For $\gamma >1$ this is equal to $T_e$ in the aging regime 
($\mu_2(t')^\gamma$ goes to zero faster than $\mu_2(t')\sim T_e(t')-T$), 
while as soon as
the VFTH exponent equals one, the correspondence breaks down and only
the asymptotic limits of the two effective temperature
will be equal to each other (and to the heat bath temperature).

\subsection{Low temperature case: $T<T_0$, $\gamma >1$}

Our approach also allows us to study the regime 
below the Kauzmann temperature 
In this last case, though, we have qualitatively
different behaviors depending on the 
value of $\gamma$, i.e., on the relative weight of $\mu_1$ and $\mu_2$.
We describe here the case $\gamma > 1$, where $\mu_1 \ll \mu_2$
\cite{LN1}.

We find  the solutions of the equations 
for the two-time correlation functions
with the following expressions for the function $\tilde{f}$:
\BEQ
{\tilde{f}}=-4 \Upsilon \Gamma(1+Q D)
-8\Upsilon\frac{Q D P}{1+Q D }(1-3r+2r^2)
+O(\mu_2\Upsilon)
\EEQ
where $r$ is an abbreviation for  the normalized difference
between the effective temperature (\ref{Te}) and the heat bath temperature
\BEQ
r\equiv \frac{T_e-T}{2T_e-T}\ .
\EEQ

The two-time correlation function turns out to be:
\BEQ
 C_{11}(t,t')\simeq\frac{1}{1+Q(t') D}
\left[m_0+\mu_2(t')
 +O(\mu_2^\gamma(t'))\right]\frac{\tilde{h}(t')}{\tilde{h}(t)}
\EEQ

For  the response function, we have
\BEQ
\fl G_{11}(t,t')\hspace*{-1mm}\simeq \hspace*{-1mm}\left[
\frac{4\Upsilon(t')\Gamma(t')}{\Kt(t')} 
 -\frac{2\Upsilon(t')(1-2r(t'))^2}{\Kt(t')}
+8\Upsilon(t')T_e(t')\left(\hspace*{-1mm}\frac{J r(t') Q(t')}{1+Q(t')D}\hspace*{-1mm}\right)^2\hspace*{-1mm}
\right]\frac{\tilde{h}(t')}{\tilde{h}(t)}
\EEQ

It follows that 
\BEQ
T_e^{F D}(t,t')\simeq T_e(t')\left[1+O\left(\mu_2(t')^{\gamma}\right)
+O(\mu_2(t')^{1+\gamma})\right].
\EEQ
In this case $O(\mu_2^\gamma)$ is always smaller than $O(\mu_2)$:
 in the long time regime $T_e^{FD}(t)$ coincides with $T_e(t)$.

With this outcome we demonstrate that it is possible to
identify an effective temperature that, coupled to the 
configurational entropy, is able to  map the dynamics of 
a system  out of equilibrium into a thermodynamic frame.
In our model, either above and below the Kauzmann transition,
such a construction seems to be 
 well founded, provided that the VFTH exponent 
$\gamma$ is bigger than one.
In this case the time evolution is so slow that a quasi-equilibrium occur
at a time dependent effective temperature. 
For $\gamma$ below one, instead, the time evolution is too fast
to allow for such a straightforward translation into  thermodynamics.

\section{Conclusions}
In the present work we have been studying a particular model glass,
that has all the basic attributes of a real glass and the 
dynamics of which can be analytically solved.

An important  assumption for our study has been the decoupling
of time-scales of the processes taking place
in the glassy dynamics.
The decoupling of time scales is also fundamental  for 
a generalization of equilibrium thermodynamics to systems 
far from equilibrium.

From the temporal behavior of the slowly varying observables in the aging 
regime we found  a VFTH relaxation time above the
 Kauzmann transition and we derived the Adam-Gibbs relation 
between the relaxation time and the configurational entropy,
which  can  be explicitly  computed for our model.
It is also possible to study the dynamics of the system 
quenched to a temperature below the Kauzmann temperature. 
For such an analysis we refer to \cite{LN1}.

We proposed an out of equilibrium thermodynamic formulation and 
we tested on our exactly solvable model whether or not such a  
generalized  approach holds, having one extra variable, namely the
{\em effective temperature}, for the
 description of  the non-equilibrium thermodynamics.
By effective temperature we mean a thermodynamic
quantity that would be the temperature  of a system at equilibrium
visiting with the same frequency the same states 
that the real - out of equilibrium - system at temperature $T$
 visits  on a given time-scale during its dynamics.
This kind of parameter appears in the thermodynamic functions
together with the heat-bath temperature and the fields coupled to
the system's observables and is coupled to the configurational entropy. 

Generally speaking,  in order to recast the out of equilibrium dynamics
into a thermodynamic frame,
the history of a system that is far from equilibrium can be 
expressed by more  than one effective parameter.
This happens when more than one  long time-scale is involved in the dynamic 
evolution of a system. In those cases to every time-sector there will
correspond an effective temperature \cite{CKL}.
Moreover, in a given time-sector,  the number of effective parameters 
needed to make such a translation into a thermodynamic frame can, 
in principle, be equal to the number of relevant observables  
considered. In our model, however,
for  certain dynamic regimes  determined by the temperature and by
the VFTH exponent $\gamma$, the effective parameters  pertaining to  
processes having the same time scale become
equal to each other for large times.
When the distance $\mu_1(t)$ from equilibrium 
is much smaller than the distance from the constraint, $\mu_2(t)$,
a single effective temperature alone is
enough for a complete thermodynamic description of the dominant
physic phenomena; this happens when $\gamma>1$.

In many models the concept of effective temperature is used in an
attempt to give a thermodynamic description of the glassy phase.
There are several analytical and numerical works defining 
the effective temperature as the fluctuation dissipation ratio
(among others \cite{DiLAPR,AFR,sFfR})
and also glassy models in which the effective temperature can be measured 
coupling the system to  a slow relaxing thermometer \cite{GR,EP}.
Even in granular systems such an approach is followed with some success
\cite{BCI}.
However, as far as we know, either no comparison is made between alternative 
(equally well based) definitions of effective temperature
or, when such a comparison is carried out, very often no coincidence is found
\cite{sFfR,FS,EP}. Nevertheless, in the $p$-spin model the fluctuation-dissipation
effective temperature coincides with the thermodynamic effective temperature;
both are equal to $T/x$ where $x$ is the break point of the Parisi function~
\cite{Nehren}~\cite{NJphysA}.

This coincidence of effective temperatures for the present model (in those 
parameter regions where it does take place) appears to be connected 
to a  slow enough relaxation dynamics. 
In order to understand what could 
happen in other models a very general analysis should be done
to identify what are the minimal requirements to produce a thermalization 
of different degrees of freedom within a single time-scale
(for a contribution in this direction see, for instance, \cite{CKP}).

\ack{The  research of L. Leuzzi  is supported by FOM (The Netherlands).}


\vskip 0.5 cm
\addcontentsline{toc}{chapter}{}
\begin{thebibliography}{99}
\bibitem{BBM} A. Barrat, R. Burioni, M. Mezard, J. Phys. A: Math. Gen.
{\bf{29}} (1996)  1311.
\bibitem{BCKM} J. P. Bouchaud, L. Cugliandolo, J. Kurchan, M.Mezard, in 
{\em Spin Glasses and Random Fields}, A. P. Young, ed.  (World Scientific, 
Singapore, 1998), p. 161.
\bibitem{VFTH}  H. Vogel,  Physik. Z. {\bf 22} (1921) 645.
 G.S. Fulcher, J. Am. Ceram. Soc. {\bf 8} (1925) 339.
 G. Tammann and G. Hesse, Z. Anorg. Allgem. Chem. {\bf 156} (1926) 245.
\bibitem{AG} G. Adam and J.H. Gibbs,
J. Chem. Phys. {\bf 43} (1965) 139.
\bibitem{KAUZ} W. Kauzmann, Chem. Rev. {\bf 43} (1948) 219.
\bibitem{CK93}
L. F. Cugliandolo, J. Kurchan, Phys. Rev. Lett. {\bf {71}} (1993) 173.
\bibitem{TOOL}
A. Q. Tool, J. Am. Ceram. Soc. {\bf{29}} (1946) 240. 
\bibitem{SKT}
F. Sciortino, W. Kob, P. Tartaglia, Phys. Rev. Lett. {\bf{83}} (1999) 3214;
W. Kob, F. Sciortino, P. Tartaglia, Europhys. Lett. {\bf{49}} (1999) 590.
\bibitem{CR}
A.Crisanti, F.Ritort, Physica A {\bf 280} (2000) 155;
  Europhys. Lett. {\bf{51}} (2000) 147.
\bibitem{N00} Th.M. Nieuwenhuizen, Phys. Rev. E {\bf{61}} (2000) 267
\bibitem{N2} Th.M. Nieuwenhuizen, J. Phys. A {\bf 31} (1998) L201;
 Phys. Rev. Lett. {\bf 79} (1997) 1317.
\bibitem{FV}
S. Franz, M. A. Virasoro, J. Phys. A {\bf{33}} (2000) 891.
\bibitem{fictive}
G.W. Scherer, {\em Relaxation in Glass and Composites} (Wiley, New York,
1986).
 S.A. Brawer, {\em Relaxation in Viscous Liquids and Glasses} (American
Ceramic Society, Columbus, OH, 1985).
 G.W. Scherer, J. Non-Cryst. Solids {\bf 123}, 75 (1990).
 A. Prados, J. J. Brey, e-print cond-mat/0103325.
\bibitem{ANGELL} C.A. Angell, Science {\bf 267} (1995) 1924.
\bibitem{BPR} L.L. Bonilla, F.G. Padilla, and F. Ritort,
Physica A {\bf 250 }(1998) 315.
\bibitem{NPRL98} Th.M. Nieuwenhuizen, Phys. Rev. Lett. {\bf 80} (1998) 5580.
\bibitem{LN2}  L. Leuzzi, T.M. Nieuwenhuizen, e-print cond-mat/0103147.
\bibitem{GR} A. Garriga, F. Ritort, Eur. Phys. J. B {\bf{ 20}} (2001) 105.
\bibitem{LN1}
L. Leuzzi, T.M. Nieuwenhuizen, Phys. Rev. E {\bf{64}} 011508 (2001).
\bibitem{NCM} Th.M. Nieuwenhuizen,  e-print cond-mat/9911052. 
\bibitem{KA} W. Kob, H.C. Andersen, Phys. Rev. E {\bf{47}} (1993) 3281.
\bibitem{BPPR} L.L. Bonilla, F.G. Padilla, G. Parisi, F. Ritort, Phys. Rev. B 
{\bf 54} (1996) 4170.
\bibitem{GLAUBER} R. J. Glauber, J. Math. Phys., {\bf{4}} (1963) 294.
\bibitem{FA84}
G.H. Fredrickson and H.C. Andersen, Phys. Rev. Lett {\bf{53}} (1984) 1244;
 J. Jackle and S. Eisinger Z. Phys. B {\bf{84}} (1991) 115;
M. Schulz, S. Trimper, Phys. Rev. B {\bf{53}} (1996) 8421.
\bibitem{FR} E. Follana, F. Ritort, 
Phys. Rev. B, {\bf 54} (1996) 930.
\bibitem{KPS} J. Kurchan, L. Peliti, M. Sellitto, 
Europhys. Lett. {\bf 39} (1997) 365.
\bibitem{KTW89} T.R. Kirkpatrick, D. Thirumalai, P.G. Wolynes, Phys. Rev. A
{\bf 40} (1989) 1045.
\bibitem{XW00} X. Xia and P.G. Wolynes, Proc. Natl. Acad. Sci. {\bf 97}
(2000) 2990.
\bibitem{KW87} T.R. Kirkpatrick and P.G. Wolynes, Phys. Rev. B {\bf 36}
(1987) 8552.
\bibitem{ParisiVF} G. Parisi, 
in {\it The Oscar Klein Centenary}, U. Lindstr\"om ed.,
(World Scientific, Singapore, 1995); e-print cond-mat/9411115.
\bibitem{CKL}
L. Cugliandolo, J. Kurchan, P. Le Doussal, 
Phys. Rev. Lett. {\bf{76}} (1996) 2390.
\bibitem{DiLAPR}
R. di Leonardo, L. Angelani, G. Parisi and G. Ruocco, 
Phys. Rew. Lett. {\bf{84}} (2000) 6054.
\bibitem{AFR}
D. Alvarez, S. Franz and F. Ritort, Phys. Rew. B {\bf{54}} (1996) 9756.
\bibitem{sFfR}
S. Franz and F. Ritort, J. Phys. A {\bf{30}} (1997) L359
\bibitem{EP}
R. Exartier and L. Peliti Eur. Phys. J. {\bf{16}} (2000) 119.
\bibitem{BCI}
L. Berthier, L.F. Cugliandolo and J.L. Iguain, Phys. Rew. E {\bf 63} (2001) 051302.
\bibitem{FS} S. Fielding and P. Sollich, e-print cond-mat/0107627.
\bibitem{Nehren}
Th. M. Nieuwenhuizen, Phys. Rew. Lett. {\bf{ 79}} (1997) 1317. 
\bibitem{NJphysA}
 Th.M. Nieuwenhuizen, J. Phys. A {\bf{31}} (1998)  L201.
\bibitem{CKP}
L.F. Cugliandolo, J. Kurchan and L. Peliti, Phys. Rew. E {\bf{55}} (1997) 3898.


\end{thebibliography}
\end{document}